\documentclass[preprint,aps,amsmath,superscriptaddress,nofootinbib,tightenlines,showpacs]{revtex4}
\usepackage{mathrsfs}
\usepackage{amsmath}
\usepackage{amsfonts}
\usepackage{amssymb}
\usepackage{latexsym}
\usepackage{graphicx}
\usepackage{bm}
\usepackage{epsfig}
\usepackage[english]{babel}
\def \be {\begin{equation}}
\def \ee {\end{equation}}
\def \bea {\begin{eqnarray}}
\def \eea {\end{eqnarray}}
\def \nn {\nonumber}

\def \a {\alpha}
\def \b {\beta}
\def \g {\gamma}

\def \d {\delta}

\def \m {\mu}
\def \n {\nu}
\def \k {\kappa}

\def \s {\sigma}
\def \r {\rho}
\def \o {\omega}

\def \th {\theta}
\def \Th {\Theta}

\def \t {\tau}
\def \dag {\dagger}
\def \p {\partial}

\def\bd{\begin{document}}
\def\ed{\end{document}}
\def\nn{\nonumber}
\def\bea{\begin{eqnarray}}
\def\eea{\end{eqnarray}}
\let\bm=\bibitem
\let\la=\label

\def\N{{\cal N}}
\def\sst{\scriptscriptstyle}
\def\thetabar{\bar\theta}
\def\Tr{{\rm Tr}}
\def\one{\mbox{1 \kern-.59em {\rm l}}}

\def\a{\alpha}      \def\da{{\dot\alpha}}
\def\b{\beta}       \def\db{{\dot\beta}}
\def\c{\gamma}  \def\C{\Gamma}  \def\cdt{\dot\gamma}
\def\d{\delta}  \def\D{\Delta}  \def\ddt{\dot\delta}
\def\e{\epsilon}        \def\vare{\varepsilon}
\def\f{\phi}    \def\F{\Phi}    \def\vvf{\f}
\def\h{\eta}
\def\k{\kappa}
\def\l{\lambda} \def\L{\Lambda}
\def\m{\mu} \def\n{\nu}
\def\o{\omega}
\def\P{\Pi}
\def\r{\rho}
\def\s{\sigma}  \def\S{\Sigma}
\def\t{\tau}
\def\th{\theta} \def\Th{\Theta} \def\vth{\vartheta}
\def\X{\Xeta}
\def\z{\zeta}
\def\w{\wedge}
\def\u{\underline}
\def\hs{\hspace}

\def\cA{{\cal A}} \def\cB{{\cal B}} \def\cC{{\cal C}}
\def\cD{{\cal D}} \def\cE{{\cal E}} \def\cF{{\cal F}}
\def\cG{{\cal G}} \def\cH{{\cal H}} \def\cI{{\cal I}}
\def\cJ{{\cal J}} \def\cK{{\cal K}} \def\cL{{\cal L}}
\def\cM{{\cal M}} \def\cN{{\cal N}} \def\cO{{\cal O}}
\def\cP{{\cal P}} \def\cQ{{\cal Q}} \def\cR{{\cal R}}
\def\cS{{\cal S}} \def\cT{{\cal T}} \def\cU{{\cal U}}
\def\cV{{\cal V}} \def\cW{{\cal W}} \def\cX{{\cal X}}
\def\cY{{\cal Y}} \def\cZ{{\cal Z}}

\def\ua{\underline{\alpha}} \def\ubb{\underline{\beta}}
\def\ug{\underline{\gamma}}
\def\ub{\underline{\phantom{\alpha}}\!\!\!\beta}
\def\uc{\underline{\phantom{\alpha}}\!\!\!\gamma}
\def\um{\underline{\mu}} \def\un{\underline{\nu}}
\def\ud{\underline\delta}
\def\ue{\underline\epsilon}
\def\una{\underline a}\def\unA{\underline A}
\def\unb{\underline b}\def\unB{\underline B}
\def\unc{\underline c}\def\unC{\underline C}
\def\und{\underline d}\def\unD{\underline D}
\def\une{\underline e}\def\unE{\underline E}
\def\unf{\underline{\phantom{e}}\!\!\!\! f}\def\unF{\underline F}
\def\unm{\underline m}\def\unM{\underline M}
\def\unn{\underline n}\def\unN{\underline N}
\def\unp{\underline{\phantom{a}}\!\!\! p}\def\unP{\underline P}
\def\unq{\underline{\phantom{a}}\!\!\! q}
\def\unQ{\underline{\phantom{A}}\!\!\!\! Q}
\def\unH{\underline{H}}
\def\ul{\underline}

\def\As {{A \hspace{-6.4pt} \slash}\;}
\def\bs {{b \hspace{-6.4pt} \slash}\;}
\def\Ds {{D \hspace{-6.4pt} \slash}\;}
\def\ds {{\del \hspace{-6.4pt} \slash}\;}
\def\ss {{\s \hspace{-6.4pt} \slash}\;}
\def\ks {{ k \hspace{-6.4pt} \slash}\;}
\def\ps {{p \hspace{-6.4pt} \slash}\;}
\def\pas {{{p_1} \hspace{-6.4pt} \slash}\;}
\def\pbs {{{p_2} \hspace{-6.4pt} \slash}\;}

\def\Fh{\hat{F}}
\def\Vh{\hat{V}}
\def\Xh{\hat{X}}
\def\ah{\hat{a}}
\def\xh{\hat{x}}
\def\yh{\hat{y}}
\def\ph{\hat{p}}
\def\xih{\hat{\xi}}

\def\psit{\tilde{\psi}}
\def\Psit{\tilde{\Psi}}
\def\tht{\tilde{\th}}

\def\At{\tilde{A}}
\def\Qt{\tilde{Q}}
\def\Rt{\tilde{R}}
\def\Nt{\tilde{N}}

\def\at{\tilde{a}}
\def\st{\tilde{s}}
\def\ft{\tilde{f}}
\def\pt{\tilde{p}}
\def\qt{\tilde{q}}
\def\vt{\tilde{v}}
\def\nt{\tilde{n}}

\def\delb{\bar{\partial}}
\def\bz{\bar{z}}
\def\bD{\bar{D}}
\def\bB{\bar{B}}

\def\bk{{\bf k}}
\def\bl{{\bf l}}
\def\bp{{\bf p}}
\def\bq{{\bf q}}
\def\br{{\bf r}}
\def\bx{{\bf x}}
\def\by{{\bf y}}
\def\bR{{\bf R}}
\def\bV{{\bf V}}

\def\d{\delta}\def\D{\Delta}\def\ddt{\dot\delta}

\def\p{\partial} \def\del{\partial}
\def\xx{\times}
\def\uno{\mbox{1 \kern-.59em {\rm l}}}

\def\trp{^{\top}}
\def\inv{^{-1}}
\def\dag{{^{\dagger}}}
\def\pr{\prime}

\def\rar{\rightarrow}
\def\lar{\leftarrow}
\def\lrar{\leftrightarrow}

\def\cw{{\cal W}}
\def\cz{{\cal Z}}
\def\tcm{\tilde{\cal M}}
\def\sgn{{\rm sgn}}
\def\sd {d^{4|4}}
\def\lan{\langle}
\def\ran{\rangle}

\def\tr{\mbox{tr}}
\def\sign{\mbox{sign}}
\def\fnl{f_\text{NL}}
\def\horava{Ho\v{r}ava}
\def\la{\langle}
\def\ra{\rangle}
\def\mb{\mathbf}
\def\nn{\nonumber}
\def\hl{Ho\v{r}ava-Lifshitz}
\def\p{\partial}
\def\dij{\delta_{ij}}
\def\tr{\mbox{tr}}
\def\sign{\mbox{sign}}
\def\fnl{f_\text{NL}}
\def\horava{Ho\v{r}ava}
\def\la{\langle}
\def\ra{\rangle}
\def\mb{\mathbf}
\def\nn{\nonumber}
\def\hl{Ho\v{r}ava-Lifshitz}
\def\p{\partial}
\def\dij{\delta_{ij}}
\def\ii{\text{i}}

\begin{document}


\title{Power spectra of scalar and tensor modes in modified \horava-Lifshitz gravity}
\author{Bin Chen\footnote{Electronic address: bchen01@pku.edu.cn},\hspace{2ex}Shi Pi\footnote{Electronic address: spi@pku.edu.cn}
and Jin-Zhang Tang\footnote{Electronic address:
JinzhangTang@pku.edu.cn}} \affiliation{Department of Physics, and
State Key Laboratory of Nuclear Physics and Technology, Peking
University, Beijing 100871, China}

\date{\today\\ \vspace{1cm}}
\begin{abstract}
In \hl~gravity, the extra dynamical scalar mode could play
significant role in cosmology. However it has been pointed out that
such a scalar may suffer from the strong coupling problem in IR. We
address this issue in this paper. Our analysis shows that the scalar
mode could decouple naturally at $\lambda=1$ due to the extra gauge
symmetry. On the other hand, the fact that the scalar mode becomes
ghost when $1/3< \lambda < 1$ is a real challenge to the theory. We
try to overcome this problem by modifying the action such that the
RG flow lies outside the problematic region. We discuss the
cosmological implications of the action and calculate the power
spectra of scalar and tensor modes. \pacs{98.80.Cq}
\end{abstract}
\maketitle

\section{introduction}\label{sec-intro}

Diffeomorphism is essential to Einstein's relativity theory of
gravity. It has been widely believed to be exact in any theory of
gravity. However, in the recent proposal by
\horava\cite{Horava:2008ih,Horava:2009uw} on gravity theory, it is
no longer an exact symmetry. The basic idea behind \horava's theory
is that time and space may have different dynamical scaling in UV
limit. This was inspired by the development in quantum critical
phenomena in condensed matter physics, with the typical model being
Lifshitz scalar field theory\cite{Lifshitz,Chen:2009ka}. In this
\horava-Lifshitz theory, the time and space will take different
scaling behavior as
\begin{equation}\label{scaling}
    \mb{x}\rightarrow b\mb{x},\;\;\;\; t\rightarrow b^zt,
\end{equation}
where $z$ is the dynamical critical exponent characterizing the
anisotropy between space and time. Due to the anisotropy, instead
of  diffeomorphism, we have the so-called foliation-preserving
diffeomorphism. The transformation is now just
 \bea
 t&\rightarrow& \tilde{t}(t) \nn\\
 x^i &\rightarrow& \tilde{x^i}(x^j,t).
 \eea

As the result of this ``reduced" gauge symmetry, there are more
physical degrees of freedom in the theory. In fact, there exist an
extra dynamical scalar degree of freedom in \hl~ gravity, as shown
in \cite{Horava:2008ih,Horava:2009uw}. This scalar degree of freedom
and its physical implication has also been discussed in cosmology
\cite{Cai:2009dx, Chen:2009jr,Wang:2009yz}. However, it was pointed
out in \cite{Charmousis:2009tc} that this scalar mode would be
strongly coupled to the matter in the IR fixed point $\lambda=1$, at
which one would expect the recovery of diffeomorphism. If this is
the case, it would lead to unacceptable effects in many experiments.
In the next section, we would like to show that this would not
happen.

The key point in our analysis is that we take the point of view that
the diffeomorphism is only an approximate symmetry even at IR. In
fact, as we will review shortly, it is not hard to see that even at
IR fixed point, there exist various other terms, involving spatial
higher-derivative terms, which break the diffeomorphism, even though
they should be very much suppressed. This is very different from the
case in Fierz-Pauli's massive gravity\cite{Fierz:1939ix}. In the
massive gravity theory, there exist extra physical degree of
freedom. It was a serious issue on how this degree of freedom get
decoupled in the massless limit, where the diffeomorphism is
completely recovered\cite{ArkaniHamed:2002sp,vanDam:1970vg}. In our
case, we will show manifestly that the extra scalar degree of
freedom could be decoupled without trouble, due to the existence of
extra gauge symmetry at IR fixed point rather than the complete
recovery of diffeomorphism. The breakdown of full diffeomorphism at
IR fixed point also suggest that the usual Stuckelberg trick could
not be used directly, especially taking into account of the
projectability condition.

Another issue on the scalar mode is whether it is a real physical
degree of freedom. The debate in the literature focus on if one
should choose the lapse function to be only the function of time,
or in other words, if the lapse function should be projectable.
The different choice seems lead to completely different physics.
For example, it was found that without the projectability
condition there were new static spherically symmetric solutions to
\hl~gravity and its modifications\cite{Lu2009, Kehagias:2009is}.
These new solutions may have profound physical implications in
solar system tests\cite{Solar}. However,  it was proved in
\cite{Tang:2009bu} that these new solutions do not respect the
projectability condition. From our point of view, taking the lapse
function as the function of time is the most natural choice. With
this choice, the gauge transformation looks transparent and
simpler. Moreover the Hamiltonian constraints form a closed
algebra, and the theory gets rid of the pathology found in
\cite{Horava:2008ih,Li:2009bg}. Simply speaking, the theory is
well-defined with the projectability condition.  As a result, the
extra scalar mode becomes the physical one and the key ingredient
in our following discussion.

On the other hand, this extra scalar mode could not be always
physical. When the parameter $\lambda$ lies between $1/3$ and $1$,
this mode is actually a ghost. We would like to emphasize that the
existence of the ghost is the real challenge to the original \hl~
gravity theory. It indicates that the theory may neither
well-defined at UV, nor UV complete. In particular, at UV, one wish
that the theory becomes non-relativistic and the speed of light is
much larger than the constant one at IR. In the original proposal of
\hl~gravity, this requires that the theory stay near $\lambda=1/3$.
However in IR, one may expect that the theory would flow to
$\lambda=1$. As the RG flow  is from $\lambda=1/3$ to $\lambda=1$,
the theory inevitably suffers from the existence of ghost. To get
away from this trouble, we try to modify the action in a way that
the RG flow may be from UV with $\lambda > 1$ to IR with
$\lambda=1$. To simplify the analysis, instead of considering the
most general form of the action, we only consider the potential with
the marginal terms and the most relevant terms. This will be the
topic in section 3.

In the remaining part of this paper, we discuss the cosmological
implications of the scalar mode, and also calculate the power
spectra of scalar and tensor perturbations. The cosmology of the
\hl~gravity has first been discussed in
\cite{Calcagni:2009ar,Kiritsis:2009sh,Lu2009,Mukohyama:2009gg}, and
then widely studied in the literature\cite{HL} from various angles.
In this paper, we take the extra scalar as an alternative to the
inflaton and study its power spectrum. We also calculate the power
spectrum of tensor mode in modified \hl~gravity action proposed in
section 3. In our treatment, we simply ignore the RG flow and use
the standard technology in cosmological perturbation theory. The
problem turns out to be quite similar to the trans-Planckian problem
in inflation. Instead of the WKB approximation used in
\cite{Chen:2009jr,Mukohyama:2009gg,Yamamoto:2009tf}, we apply the
technology in trans-Planckian physics and study the equation of
motion of scalar perturbation stage by stage. We find that the power
spectra are scalar invariant. This is not a surprise since the
classical evolution is a pure de-Sitter phase, which has time
translation invariance. We also notice that the tensor-to-scalar
ratio is sensitive to the time of horizon-crossing of tensor and
scalar modes, and can be small if at the time of scalar crossing the
horizon $\l$ is near $1$.

The paper is organized as follows. In section 2, we study the
gravitational scalar in the \hl~gravity theory. In section 3, we
present our modification of the \hl~gravity action. In section 4 and
5, we calculate the power spectra of the scalar and tensor
perturbations respectively. We end with some discussions in section
6.

\section{scalar mode in \hl~gravity}

 Since time direction plays a privileged role in the whole
construction, it is more convenient to work with ADM metric
\begin{equation}\label{ADMmetric}
    ds^2=-N^2dt^2+g_{ij}(dx^i+N^idt)(dx^j+N^jdt).
\end{equation}
Due to the anisotropy between time and space, the usual
diffeomorphisms reduce to  the foliation-preserving diffeomorphisms,
generated by infinitesimal transformation:
 \be
 \delta t= \xi^0(t), ~~~\delta x^i=\xi^i(t,\vec{x}).
 \ee
 The essential point is that $\xi^0$ is just the function of $t$.
This leads to the following transformations on the metric
components:
 \bea
 \delta g_{ij}&=&\p_i\xi^kg_{jk}+\p_j\xi^k
 g_{ik}+\xi^k\p_kg_{ij}+\xi^0 \dot{g}_{ij}, \nn\\
 \delta N_i&=&\p_i\xi^jN_j+\xi^j\p_j
 N_i+\dot{\xi}^jg_{ij}+\dot{\xi}^0N_i+\xi^0\dot{N}_i, \nn\\
 \delta N&=&\xi^j\p_j N +\dot{\xi}^0N+\xi^0\dot{N}. \label{gaugetr}
 \eea
 The above transformations could be obtained by taking a
 nonrelativistic limit of usual relativistic diffeomorphisms.
 It is more convenient and natural to choose $N$ being just the function of
 $t$. There are a few advantages to work with this choice. With this choice,
 the gauge symmetry is simpler and transparent. Furthermore, in the
 Hamiltonian formulation, the constraints could form a closed
 algebra since the momentum conjugate to $N$
 does not lead to a local constraint\cite{Horava:2008ih}. As a result of less constraints than standard GR,
 the physical
 degrees of freedom in the theory include not only the massless gravitons
 but also another propagating scalar. The existence of extra
 scalar field has profound meaning in cosmology. In \cite{Chen:2009jr}, we showed that
 for the action without the detailed balance condition, this scalar may lead to scale
 invariant spectrum.

 On the other hand, if one abandon the projectability condition and
 let $N$ be the function of both $t$ and $x^i$, one will find
 that the theory would be ill-defined, as shown in
 \cite{Horava:2008ih,Li:2009bg}.

 At the special value $\lambda=1$, the theory develops an enhanced
  time-independent $U(1)$ gauge symmetry acting via
 \be
 \delta N_i=\p_i\epsilon, ~~\delta g_{ij}=0.
 \ee
 Due to the existence of extra gauge symmetry, the scalar mode is
 not physical anymore. It is remarkable that even with this extra
 gauge symmetry, the total gauge symmetries is different from the
 usual
 diffeomorphisms in general relativity. In other words, the diffeomorphisms has not been
 recovered at $\lambda=1$. This fact is essential to understand why at
 $\lambda=1$ the extra scalar degree of freedom could be decoupled
 without trouble.

To understand the decoupling better, let us consider the
perturbation around FRW metric:
 \bea
 ds^2&=&-dt^2+a(t)^2 \delta_{ij}dx^i dx^j \nn\\
 &=& a^2(\eta)(-d\eta^2+\delta_{ij}dx^i dx^j).
 \eea
Here for simplicity we focus on the flat universe, and use the
co-moving time $\eta=\int dt/a$ as time variable. The above metric
could be reduced to Minkovski spacetime if $a(t)$ is a constant.
Perturb the flat metric, and use the ADM formulism in co-moving
time,
\begin{equation}\label{ADM:co-moving}
    ds^2=-(\mathcal{N}^2-\mathcal{N}^i\mathcal{N}_i)d\eta^2+2\mathcal{N}_id\eta
    dx^i+g_{ij}dx^idx^j,
\end{equation}
The fluctuations around the above metric could be
\begin{eqnarray}\label{N}
 \mathcal{N}&=&a(\eta)(1+A), \label{Ni}\\\mathcal{N}_i&=&a(\eta)(\p_i
 B+V_i),\label{gij}
 \\g_{ij}&=&a^2(\eta)\{(1-2\psi)\delta_{ij}-\p_i\p_jE-2\p_{(i}F_{j)}+h_{ij}\},
\end{eqnarray}
where $A,B,\psi, E$ are scalar perturbations, $V_i$ and $F_j$ are
vector perturbations, and $t_{ij}$ is the gauge-invariant tensor
perturbation describing gravitational wave. Under the gauge
transformations (\ref{gaugetr}), we have
 \bea
 A&\rightarrow & \tilde{A}=A-\frac{1}{a}(\xi^0 a)^\prime, \\
 B&\rightarrow & \tilde{B}=B-a\zeta^\prime, \\
 E&\rightarrow & \tilde{E}=E+2\zeta, \\
 \psi&\rightarrow & \tilde{\psi}=\psi+\xi^0\frac{a^\prime}{a}, \\
 V_i&\rightarrow & \tilde{V}_i=V_i+(\xi_{i\bot})^\prime,\\
 F_j&\rightarrow & \tilde{F}_j=F_j+(\xi_{i\bot})^\prime, \\
 h_{ij}&\rightarrow & \tilde{h}_{ij}=h_{ij},
 \eea
where we have decompose the spatial vector $\xi^i$ as
 \be
 \xi^i=\xi^i_{\bot}+\p^i\zeta
 \ee
 with $\xi^i_{\bot}$ being divergenceless and $\zeta$ being a
 scalar. The gauge invariant variables besides $h_{ij}$ are
 \bea
 \Psi&=&A+\psi+\left(\frac{\psi}{\mathscr{H}}\right)^\prime, \\
 \Phi&=&B+\frac{E^\prime}{2}, \\
 S_i&=&V_i-F_i.
 \eea
 We will only focus on the scalar perturbations. It is convenient
 to work with the gauge
 \be\label{gauge}
 A=0, ~~~ E=0.
 \ee
Note that the above gauge choice is consistent with the
projectability condition. Since $\xi^0$ is the only function of $t$,
the gauge transformation on $A$ would not spoil the projectability
condition.

If the scale factor is a constant, the above gauge transformations
reduce to the ones in flat spacetime. In \cite{Kim:2009zn}, the
gauge invariant perturbations about the flat spacetime  have been
analyzed carefully. Actually, from the discussion there, one can
see that the extra dynamical scalar mode can decouple without
trouble. It was claimed in \cite{Kim:2009zn} that such scalar mode
is not a propagating mode. This is not true. The problem comes
from the fact that the lapse function is projectable so that it
induce a non-local super-Hamiltonian constraint. For the flat
spacetime, the perturbation $A$ of the lapse function is a pure
gauge and can be set to zero safely. Even if  $A$ is kept
nonvanishing, the variation with respect to $A$ would only lead to
non-local constraint, which is less powerful than the local one.

 In terms of ADM
metric, the action of original \hl~gravity theory can be written
as\cite{Horava:2009uw}
\begin{eqnarray}
\label{S_g:origin} \nn
    S_g=\int
    dtd^3\mb{x}\sqrt{g}N&&\left\{\frac{2}{\kappa^2}K_{ij}G^{ijkl}K_{kl}
    -\frac{\kappa^2}{2}\left[\frac{1}{\omega^2}C_{ij}-\frac{\mu}{2}\left(R_{ij}
    -\frac{1}{2}Rg_{ij}+\Lambda_Wg_{ij}\right)\right]\right.\\
    &&\cdot
    \left.{}G^{ijkl}\left[\frac{1}{\omega^2}C_{ij}
    -\frac{\mu}{2}\left(R_{ij}-\frac{1}{2}Rg_{ij}+\Lambda_Wg_{ij}\right)\right]\right\}.
\end{eqnarray}
where $K_{ij}$ is the extrinsic curvature of the spatial
hypersurface; $C_{ij}$ is the Cotton tensor which can be used to
preserve the detailed-balanced condition in constructing the action;
$G^{ijkl}$ is the De Witt metric on the space of metrics that
preserve the anisotropic diffeomorphism, and $R_{ij}$ is the Ricci
tensor in spatial hypersurface. Their definitions are
\begin{eqnarray}\label{K}
K_{ij}&=&\frac{1}{2N}(\dot{g}_{ij}-\nabla_iN_j-\nabla_jN_i),\\\label{Coton}
C_{ij}&=&\epsilon^{ikl}\nabla_k\left(R^j{}_l-\frac{1}{4}R\delta^j_l\right),\\\label{DeWitt}
G^{ijkl}&=&\frac{1}{2}\left(g^{ik}g^{jl}+g^{il}g^{kl}\right)-\lambda
g^{ij}g^{kl}.
\end{eqnarray}
Here and throughout the paper, a dot over the quantity means taking
the derivative with respect to cosmic time $t$, while a prime
denotes that to co-moving time $\eta$. The first term in
\eqref{S_g:origin} involving only the extrinsic curvature is the
kinetic term, while the others are potential terms. $\lambda$ is the
coupling constant in the kinetic term, and runs expectedly to
$\lambda=1$ at IR regime at which the kinetic term goes back to the
one in the general relativity. This specific form of the action  is
governed by the detailed-balance condition, which is just applied by
\horava for convenience to decease the number of arbitrary
parameters. The expansion of the action gives
\begin{eqnarray}\label{S_g:expand}\nn
    S_g&=&\int dtd^3\mb{x}\sqrt{g}N\left\{\frac{2}{\kappa^2}(K_{ij}K^{ij}-\lambda K^2)
    -\frac{\kappa^2}{2\omega^4}C_{ij}C^{ij}+\frac{\kappa^2\mu}{2\omega^2}\epsilon^{ijk}
    R_{il}\nabla_jR^l{}_k\right.\\
    &&\left.-\frac{\kappa^2\mu^2}{8}R_{ij}R^{ij}
    +\frac{\kappa^2\mu^2}{8(1-3\lambda)}\left(\frac{1-4\lambda}{4}R^2
    +\Lambda_WR-3\Lambda_W^2\right)\right\}.
\end{eqnarray}
Comparing this action with the Einstein-Hilbert action in IR limit
\begin{equation}\label{EHaction}
    S_{\text{EH}}=\frac{1}{16\pi G}\int
    d^4x\sqrt{g}N\{(K_{ij}K^{ij}-K^2)+R-2\Lambda\},
\end{equation}
with $x^0\equiv ct$, we can recover the speed of light, Newton
constant and the cosmological constant by the parameters introduced
before,
\begin{eqnarray}\label{c:former}
  c = \frac{\kappa^2\mu}{4}\sqrt{\frac{\Lambda_W}{1-3\lambda}},
  \hspace{3ex}
  G = \frac{\kappa^2}{32\pi c},
  \hspace{3ex}\Lambda=\frac{3}{2}\Lambda_W.
\end{eqnarray}
Thus at IR the theory  recovers nearly the usual general relativity,
with the higher derivative terms of spatial metric components as the
modifications. Even though the higher derivative terms are highly
suppressed at IR, strictly speaking, the theory always breaks
diffeomorphism, or locally Lorentz invariance.

In \horava's original paper \cite{Horava:2009uw} the coupling
constant $\lambda$ runs to 1 in IR limit. And in UV, because of
the anisotropy between space and time, the speed of light is not a
constant and may be  extremely large, which could be used to
explain the horizon and flatness problem\cite{Kiritsis:2009sh}.
But from \eqref{c:former} we know that this can only occur in the
case $\lambda<1/3$ if we take $\Lambda$ to be positive, taking
into account of the fact $\Lambda$ is directly related to
cosmological constant. However, this raise the worry that  the
marginal coupling constant $\lambda$  can never run to its
infrared value $\lambda=1$, which is directly in contrast with our
former description.
To solve this problem, it was proposed that one should do
analytical continuation on the parameters\cite{Lu2009}
\begin{equation}\label{continuation}
    \mu\rightarrow i\mu,\;\;\;\;\omega\rightarrow-i\omega,
\end{equation}
which leaves the action real.
And under this continuation, we see from \eqref{c:former} that
\begin{equation}\label{c:continuation}
  c = \frac{\kappa^2\mu}{4}\sqrt{\frac{\Lambda_W}{3\lambda-1}},
\end{equation}
and there is no conflict between $\Lambda>0$ and $\lambda>1/3$.
And when $\lambda\rightarrow1/3$ proposed by \horava~as the
ultraviolet value of this coupling constant, we have a very large
speed of light, which can naturally solve the casuality problem in
cosmology without inflation.

To understand the extra scalar degree of freedom, let us come back
to the scalar perturbation we studied
before\cite{Cai:2009dx,Chen:2009jr}. We need only focus on the
kinetic term in the above action, which is just
 \be
S_K = \int dtd^3\mb{x}\left\{3\alpha a^3(1-3\lambda)
  \left[\frac{2}{3}\frac{\dot{\psi}^2}{1-\lambda}
  +6H\psi\dot{\psi}+9H^2\psi^2\right]\right\}.
\ee Several remarks are in order:
\begin{enumerate}
 \item From the action, it is obvious that the scalar mode $\psi$
is physical when $\lambda < 1/3$ and $\lambda >1$, while when
$1/3<\lambda <1$ the mode is a ghost, indicating that the theory is
not well-defined\cite{Bogdanos:2009uj}. At the special value
$\lambda=1$, the mode is decoupled, as we will clarify more below.
And at $\lambda=1/3$, the theory has extra symmetry, as discussed
carefully in \cite{Horava:2009uw}. This fact is the same as the one
found in \cite{Horava:2008ih, Horava:2009uw} where the perturbations
around the  flat spacetime were studied.
 \item More interestingly, the equation of motion of $\psi$ takes
 the following form:
 \be
 \frac{1-3\l}{1-\l}\ddot{\psi}+....
 \ee
 This indicates that when $\l \to 1$, the scalar field $\psi$
 could be decoupled naturally, in contract with the claim in
 \cite{Charmousis:2009tc}. It seems that the strong coupling problem
 does not exist in our case.
 \item The absence of the strong coupling problem may stem from the
fact that we take different  points of view on gauge
transformations. In our case, we stick to the requirement that the
lapse function should be projectable, as originally advocated in
\cite{Horava:2009uw}. As a result, we do not expect that the
diffeomorphism is recovered at $\lambda=1$. Instead, the decoupling
of the extra scalar mode comes from the fact that there is extra
gauge symmetry at $\l=1$. This is conceptually different from the
case studied in \cite{Charmousis:2009tc} and Fierz-Pauli massive
gravity\cite{Fierz:1939ix}.
 \item Technically it is remarkable the equation of motion of
 $\psi$ has a prefactor  proportional to $1/(1-\l)$ rather
 than $(1-\l)$. This difference has significant physical
 implication. In our case, this means that the scalar mode could
 be decoupled without trouble. Another way to see this is to
 cast the scalar mode into canonical form such that the mode
 become non-physical at $\l=1$. It is remarkable that in \cite{Horava:2008ih,
 Horava:2009uw}, the equation of motion of the scalar mode around
 the  flat spacetime background has the prefactor $(1-\l)$. However  this is due to different gauge choice. It has been
 shown in \cite{Kim:2009zn} by rescaling the field, one has the same
 equation of motion. In fact, no matter what kind of gauge choice,
 the physical dispersion relation is exactly the same. This suggests
 that for the cosmological perturbations, the different gauge choice
 would not lead to different dispersion relation. Namely, the extra
 scalar mode may decouple naturally as $\l \to 1$.
\end{enumerate}

\section{Modified \hl~ gravity}

The existence of the ghost is fatal to the theory. It means that the
theory is not well-defined, not mentioning UV completeness. One may
expect that we can always work in the region outside $\l \in [1/3,
1]$. However this cannot be guaranteed, considering our ignorance of
the details of RG flow. On the other hand, in the practical
application in cosmology, one wish the RG flow is from $\l \sim 1/3$
to $\l =1$ in original \hl~ gravity. In this paper, we take a modest
attitude and try to modify the \hl~ gravity such that the RG flow
may happen always with $\l >1$. In order to do so, we have to
abandon the detailed balance condition. As it is well-known, the
detailed balance condition may not be essential to the
theory\cite{Calcagni:2009qw,Sotiriou:2009bx,Carloni:2009jc}. The
imposing of such condition is pragmatic to simplify the action. In
principle, one may relax this condition and consider more general
form of the action. In this paper, we do not want to consider the
most general form of the action. Instead, we just consider the
marginal spatial kinetic part and most relevant deformations,
besides the time kinetic terms. The action we start with is of the
form
\begin{eqnarray}\label{S:without Balance}\nn
    S_g&=&\int dtd^3\mb{x}\sqrt{g}N\left\{\alpha(K_{ij}K^{ij}-\lambda K^2)+\xi(\lambda)R+\sigma(\lambda)
\right.\\
&&\left.
-\beta\left(\beta_1\nabla_iR_{jk}\nabla^iR^{jk}+\beta_2\nabla_iR_{jk}\nabla^jR^{ik}+\beta_3\nabla_iR\nabla^iR\right)\right\}.
\end{eqnarray}
Here we only keep the marginal terms that are power-counting
renormalizable and dominant in UV limit, besides the
lower-dimensional terms to recover IR behaviors.\footnote{Actually,
after some integrals by parts and using the Bianchi identity, the
$\beta_2\nabla_iR_{jk}\nabla^jR^{ik}$ term can be converted to a
$(\beta_2/4)\nabla_iR\nabla^iR$ term and some higher order terms. In
our current work we can just set $\tilde{\beta}_3=\beta_3+\beta_2/4$
and discard the $\beta_2$ term. This will change in the calculation
of non-Gaussianity. Thanks Shinji Mukohyama for useful discussions.}
 The other marginal terms being cubic of Ricci scalar and Ricci tensor,  and the other relevant
terms like $R^2$ and $R\nabla R$ are neglected for simplicity. For
a complete discussion on all possible terms maintaining  the
power-counting renormalizability, see \cite{Sotiriou:2009bx}.

 Because of the breakdown of the detailed balance condition,
the coupling constants before each terms are independent. The
couplings could be connected to the speed of light, the Newtonian
coupling constant and the cosmological constant of Einstein's
general relativity in IR limit,
\begin{eqnarray}
\label{c}
  c^2 &=& \frac{\xi}{\alpha}, \\\label{G}
  16\pi G &=& \frac{1}{c\alpha},\\\label{Lambda}
  \Lambda&=&-\frac{\sigma}{2\xi}.
\end{eqnarray}
Here we see that $c^2$ can be positive, if we choose a proper form
of the function $\xi(\lambda)$. Furthermore, we can require $c$ to
be very large when $\lambda$ is near its ultraviolet value. In
\horava's original paper, he suggested $\lambda\rightarrow1/3$ at
the UV limit, which gives a large speed of light in \eqref{c:former}
or \eqref{c:continuation}. Here we only take this condition as a
constraint on the function $\xi(\lambda)$. For instance if the
theory requires $\lambda$ to be larger than the unity at UV as we
will propose as a condition to exclude the ghost field, the function
$\xi(\lambda)$ may be divergent when $\lambda$ tends to be infinity.

For our use let us have a glance of the classical dynamics of the
universe under such an action. In a homogenous and isotropic
universe,
\begin{equation}\label{metric:BG}
    ds^2=-dt^2+a^2h_{ij}dx^idx^j,\;\;\;\;h_{ij}=\delta_{ij}+\frac{\mathcal{K}x^ix^j}
    {1-\mathcal{K}\mb{x}^2},
\end{equation}
where $\mathcal{K}$ is the parameter to describe the spatial
curvature. Under this metric, the universe is homogeneous and
isotropic, which will greatly simplify our following calculations.
To apply our foliated diffeomorphism, we need use the ADM formalism
of this Robertson-Walker metric, with the extrinsic curvature and
the Ricci tensor to be
\begin{eqnarray}\label{K:bg}
  K_{ij}=H(t)g_{ij},& &K=3H(t), \\\label{R:bg}
  R_{ij}=\frac{2\mathcal{K}}{a^2}g_{ij}, & &
  R=\frac{6\mathcal{K}}{a^2},
\end{eqnarray}
where $H(t)=\dot{a}/a$ is the Hubble parameter.

We take the variation of the action \eqref{S:without Balance} with
respect to $N$, and have our first equation of constraint.
\begin{equation}\label{constraint:N}
    \int\sqrt{g}\left[-\frac{2}{\kappa^2}(K^{ij}K_{ij}-\lambda K^2)+\zeta
    R+\sigma\right]d^3\mb{x}=\int\sqrt{g}\rho d^3\mb{x}.
\end{equation}
Here, $\rho$ is the energy density of the Lifshitz scalar in the
universe, and can be written as
\begin{equation}\label{EnergyDensity}
    \rho=-\frac{1}{\sqrt{g}}\frac{\delta S_m}{\delta N}
\end{equation}
where $S_m$ is the action of matter field, which can be a Lifshitz
scalar\cite{Calcagni:2009ar}, gauge field\cite{Chen:2009ka} or
something else. Because of the projectability of the lapse function
$N(t)$, we only have a spatial-integral constraint here. This is
generic for all the \horava-like models with  projectable lapse
function $N(t)$\cite{Sotiriou:2009bx}. But, for a homogeneous and
isotropic Friedman universe, this constraint equation is valid at
every point, and the integral can be removed legally. Thus we have
the first Friedman's equation\cite{Lu2009,Calcagni:2009ar}
\begin{equation}\label{Friedman:H}
  H^2=\frac{c^2}{3\alpha(1-3\lambda)}\left[-\rho+\frac{6\mathcal{K}}{a^2}\xi(\lambda)
  +\sigma(\lambda)\right].
\end{equation}
Since $\rho$ is the energy density of matter and radiation,
$\sigma(\lambda)$ plays the role of ``cosmological constant". Here,
it is a function of $\lambda$ and evolves when $\lambda$ varies as
the energy scale changes. This implicant dependence may be treated
carefully when we are facing problems like the evolution of dark
energy or the tilt of the power spectrum. But because the dependence
of $\lambda$ on the cosmic time is unknown, we will neglect this
dependence and suppose that in the process we are interested in, the
change of $\sigma(\lambda)$ is so little that it will not have any
significant physical effect, and so is $H(\lambda)$.
We see from \eqref{Friedman:H} that if the universe is flat and
dominated by the cosmological constant, for some $\lambda$ greater
than 1/3, we must have $\sigma(\lambda>1/3)<0$, which means that
we have a positive cosmological constant $\Lambda>0$ at IR, since
from $\eqref{c}$, $\xi(\lambda>1/3)$ is always positive. These two
conditions guaranteed the positivity of the cosmological constant
and $H^2$. If the matter/radiation contribution could be ignored
safely, the homogenous and isotropic solution  is a pure de-Sitter
spacetime, with an exponentially expanding scale factor
$a(t)\propto\exp(Ht)$.

The second equation of constraint is obtained by taking the
variation of the action with respect to the shift vector $N_i$,
\begin{equation}\label{constraint:N_i}
    \nabla_i(K^{ij}-\lambda Kg^{ij})=0.
\end{equation}
Because the extrinsic curvature is homogeneous in a Friedman
universe, as in \eqref{K:bg}, $K_{ij}\propto g_{ij}$, this
equation is trivially satisfied for the background evolution. But
it will supply a perturbative constraint equation up to first
order if the perturbations to the background metric are under
consideration.

Finally take the variation of action \eqref{S:without Balance} with
respect to $g_{ij}$, we have the equation of motion of dynamical
degree of freedom. The explicit expression (see
\cite{Lu2009,Kiritsis:2009sh,Sotiriou:2009bx}) is rather lengthy and
has little to do with our following discussion so we would like not
write it here.

\section{spectrum of the gravitational scalar}
From the discussion above, we know that the classical evolution of
the scale factor in the \horava~era is determined by
(\ref{Friedman:H}). Especially, when cosmological constant is
dominant and the universe is flat, the evolution is the
exponentially expansion like in a de Sitter phase. In this section,
we will calculate the perturbation of the gravitational field, and
study the equations of motion of the scalar modes.

Taken the ADM formalism and the gauge choice (\ref{gauge}), with
some relations derived through the two constraint equations, the
perturbed action of the gravitational field up to second order can
be written as
\begin{eqnarray}\label{Action_2}\nn
  S^{(2)} &=& \int dtd^3\mb{x}\left\{3\alpha a^3(1-3\lambda)
  \left[\frac{2}{3}\frac{\dot{\psi}^2}{1-\lambda}
  +6H\psi\dot{\psi}+9H^2\psi^2\right]\right. \\
   &&\left.-\frac{2\beta}{a^3}(3\beta_1+2\beta_2+8\beta_3)\psi\p^6\psi-2a\xi
   (\lambda)\psi\p^2\psi\right\}.
\end{eqnarray}
Now the Hubble parameter is a constant. For convenience we define a
conformal time $\eta$ with $dt=ad\eta$ and introduce an auxiliary
field $\chi=a\psi$. After taking the variation with respect to
$\psi$, and changing to the momentum space, we have
\begin{equation}\label{eom:chi-simple}
    \chi''(\eta)+\left(k^6H^4\bar{L}^4\eta^4+c_s^2k^2-\frac{2}{\eta^2}\right)\chi(\eta)=0.
\end{equation}
where
\begin{equation}\label{cs}
    c_s^2=\frac{1-\lambda}{1-3\lambda}c^2
\end{equation}
is the speed of sound, and
\begin{equation}\label{L}
    \bar{L}=\frac{L}{2\pi}, \hspace{3ex}L=2\pi\left[\frac{\beta}{\alpha}\frac{1-\lambda}{1-3\lambda}(3\beta_1+2\beta_2+8\beta_3)\right]^\frac{1}{4},
\end{equation}
is the characteristic length which denotes the scale where the
trans-Planckian effects becomes significant.

This equation can not be solved analytically. However, many efforts
has been done to deal with this type of Corley-Jacobson dispersion
relation\cite{Martin:2000xs,Martin:2002kt,Brandenberger:2000wr}. The
author of \cite{Koh:2009cy} have already studied the trans-Planckian
physics appearing naturally in Horava-Lifshitz gravitational waves,
and find a scale invariant power spectrum in a specific connecting
time. Here we follow the method of Martin et.al.\cite{Martin:2000xs}
to investigate the trans-Planckian effects of the gravitational
scalar. We split the period under consideration into three regions
by two different characteristic lengths: the trans-Planckian length
$L$ and the sound horizon $c_s/H$. First, when
$\eta\rightarrow-\infty$, we can set the initial value of the wave
function $\chi(\eta_\ii)$ and its derivative $\chi'(\eta_\ii)$ such
that they initially minimize the energy density. This is to satisfy
\begin{eqnarray}
  \chi(\eta_\ii) &=& \frac{1}{\sqrt{2k^3}}\frac{1}{H\bar{L}|\eta_\ii|}, \\
  \chi'(\eta_\ii) &=& \pm i\sqrt{\frac{k^3}{2}}H\bar{L}|\eta_\ii|.
\end{eqnarray}
Then, in UV region when the physical wavelength of the mode
concerned is much smaller then the characteristic length, $k^6$ term
is dominant in \eqref{eom:chi-simple},
\begin{equation}\label{eom:I}
    \chi_\mathrm{UV}''+k^6H^4\bar{L}^4\eta^4\chi_\mathrm{UV}=0.
\end{equation}
In a new variable $z=k^3H^2\bar{L}^2|\eta|^3/3$,  the solution to
this equation can be expressed as
\begin{equation}\label{chi_UV}
\chi_\mathrm{UV}(\eta)=A_1\sqrt{|\eta|}J_{1/6}(z)+A_2\sqrt{|\eta|}J_{-1/6}(z),
\end{equation}
where $A_1$ and $A_2$ can be determined by the continuity of $\chi$
and $\chi'$ at the initial time $\eta_i$, to be
\begin{eqnarray}
  A_1 &=& \frac{\pi H\bar{L}}{6\sin(\pi/6)}\sqrt{\frac{k^3|\eta_\ii|^3}{2}}[J_{5/6}(z_\ii)\mp iJ_{-1/6}(z_\ii)], \\
  A_2 &=& \frac{\pi H\bar{L}}{6\sin(\pi/6)}\sqrt{\frac{k^3|\eta_\ii|^3}{2}}[J_{-5/6}(z_i)\mp iJ_{1/6}(z_\ii)],
\end{eqnarray}
Note, that in the UV region $|\eta_\ii|\gg1$ and $z_\ii\gg1$, we can
expand the Bessel's function into its asymptotic form when the
argument is large, and the coefficients reads
\begin{eqnarray}
  A_1 &\approx& \sqrt{\frac{\pi}{3}}(\mp i)\exp\left[\mp i\left(z_\ii+\frac{\pi}{12}-\frac{\pi}{4}\right)\right]
  \equiv\pm i\sqrt{\frac{\pi}{3}}e^{\mp ix_\ii},\\
  A_2 &\approx& \sqrt{\frac{\pi}{3}}(\pm i)\exp\left[\pm i\left(z_\ii-\frac{\pi}{12}-\frac{\pi}{4}\right)\right]
  \equiv\pm i\sqrt{\frac{\pi}{3}}e^{\pm iy_\ii}.
\end{eqnarray}
Here for simplicity we have defined
\begin{equation}\label{def:x&y}
    x=z+\frac{\pi}{12}-\frac{\pi}{4},\;\;\;\;y=z-\frac{\pi}{12}-\frac{\pi}{4}.
\end{equation}

In the intermediate region, the wavelength of the $k$-mode exceeds
the characteristic length $L$ but still much less then the sound
horizon $c_s/H$, i. e. $L\gg\lambda\gg c_sH^{-1}$. Then we can
neglect the UV term and the cosmological damping term in
\eqref{eom:chi-simple}, and get an oscillation solution of the
perturbation as a plane wave. We can deduce the connecting time
$\eta_\star$ between UV and intermediate region \footnote{Here some
subtlety exists when choosing the exact connecting time whether we
should let $\lambda=L$ or $k^6H^4\bar{L}^4\eta^4=c_s^2k^2$. In
\cite{Martin:2002kt}, the detailed discussion shows that we should
choose to avoid the oscillation spectrum. But in our case we will
show that the former choice will also sweep the oscillation in the
final result.},
\begin{equation}\label{eta1}
    \frac{2\pi}{k}a(\eta_\star)=L_\star,\;\;\;\;|\eta_\star|=\frac{2\pi}{kHL_\star}=\frac{1}{kH\bar{L}_\star}.
\end{equation}
Here, the emergence of the subscript $\star$ of the characteristic
length $L_\star$ is because that  $L$ also varies with time through
the parameter $\lambda$, which depends on the energy scale by the
renormalization flow. When we calculate the critical time
$\eta_\star$, all the parameters should be the value at $\eta_\star$
including $L$. In this region, the solution is the usual plane wave
\begin{equation}\label{chi_inter}
\chi_\mathrm{int}(\eta)=B_1e^{ic_sk\eta}+B_2e^{-ic_sk\eta}.
\end{equation}
Then we can determine the coefficients $B_1$ and $B_2$ by connecting
$\chi_\mathrm{int}$ and its derivative with those in the UV region.
This reads
\begin{eqnarray}
\nonumber
  B_1e^{ic_sk\eta_\star}&=&\frac{A_1}{2}|\eta_\star|^{1/2}\left[J_{1/6}(z_\star)+ik^2H^2\bar{L}^2|\eta_\star|^2J_{-5/6}(z_\star)\right] \\
   & & +\frac{A_2}{2}|\eta_\star|^{1/2}\left[J_{-1/6}(z_\star)-ik^2H^2\bar{L}^2|\eta_\star|^2J_{5/6}(z_\star)\right], \\
B_2e^{-ic_sk\eta_\star}&=&\frac{A_1}{2}|\eta_\star|^{1/2}\left[J_{1/6}(z_\star)-ik^2H^2\bar{L}^2|\eta_\star|^2J_{-5/6}(z_\star)\right] \nn \\
   & &+\frac{A_2}{2}|\eta_\star|^{1/2}\left[J_{-1/6}(z_\star)+ik^2H^2\bar{L}^2|\eta_\star|^2J_{5/6}(z_\star)\right].
\end{eqnarray}
To go further we notice that
\begin{equation}
k^2H^2\bar{L}_\star^2|\eta_\star|^2=1,
\end{equation}
and generally $L_\star\ll H^{-1}$, so we also have
$|\eta_\ii|\gg|\eta_\star|\gg1$. The Bessel functions can also be
expanded in its asymptotic form as before, then we get
\begin{eqnarray}
  B_1 &=& \mp\frac{i}{\sqrt{2k}}e^{-ic_sk\eta_\star}e^{\pm ix_\ii}[e^{-iy_\star}-e^{\mp i\pi/6}e^{-ix_\star}], \\
  B_2 &=& \mp\frac{i}{\sqrt{2k}}e^{ic_sk\eta_\star}e^{\pm ix_\ii}[e^{iy_\star}-e^{\mp
  i\pi/6}e^{ix_\star}],
\end{eqnarray}
where $x$ and $y$ are defined in \eqref{def:x&y}.

As the universe expands the wavelength of $k$-mode is stretched and
becomes larger and larger, and finally exceeds the sound horizon.
This critical time $\eta_\ast$ is determined by
\begin{equation}
    \frac{a(\eta_\ast)}{k}=\frac{c_{s\ast}}{H},\;\;\;\;|\eta_\ast|=\frac{1}{c_{s\ast}k}.
\end{equation}
The meaning of the subscript $\ast$ of the sound speed is similar as
before: we require $\lambda$ in the definition of $c_s$ be its value
at $\eta_\ast$. When $\eta\gg\eta_\ast$, the fluctuation is in an IR
region. The perturbation will freeze out after it exceeds the
horizon, so primordial value of the power spectrum observed today
can be traced back to its value at and before $\eta_\ast$. The IR
solution of equation \eqref{eom:chi-simple} is
\begin{equation}
    \chi_\mathrm{IR}=C\eta^2+\frac{D}{\eta}
\end{equation}
For convenience we only pick the increasing mode\footnote{Connect
with both the decaying and increasing modes will bring here an
inessential factor of order unity, as in \cite{koh}.}, whose
coefficient can be determined by connecting $\chi_\mathrm{int}$ and
$\chi_\mathrm{IR}$ at $\eta_\ast$, which reads
\begin{equation}
    D=\frac{1}{c_{s\ast}\sqrt{2k^3}}\exp\left\{\mp
    i\left[k(\eta_\star-\eta_\ast)+z_\star-z_\ii-\frac{\pi}{2}\right]\right\}.
\end{equation}
Therefore the power spectrum is
\begin{equation}\label{spectrum:scalar}
    \mathscr{P}_\psi=\frac{k^3}{2\pi^2}|\psi|^2=\frac{k^3}{2\pi^2}H^2|D|^2
    =\left(\frac{H}{2\pi c_{s}}\right)_\ast^2.
\end{equation}
which is obviously scale-invariant, if we neglect the time variation
of the horizon at the inflationary stage. If so, the slight
difference of the horizon-crossing time for different wavelengths
will produce different $H$'s, thus different spectra. This is of
course the usual case in the model slightly breaking the
time-translation invariance of the de-Sitter stage.

To comprehend the significance of \eqref{spectrum:scalar}, we note
that the spectrum is frozen out after the mode exceeds the sound
horizon. When the exponentially expansion is over and the universe
recovers usual GR behavior, all the parameters in our original
action \eqref{S:without Balance} recover the IR limit value, and
specifically,  $\lambda\rightarrow1^+$. But this will never
influence the value of the power spectrum which is completely
determined by the values of parameters at horizon-crossing. On the
other hand, if $\lambda$ runs to unity much earlier before the
wavelength of the fluctuation mode exceeds the sound horizon, i. e.
$c_s(\eta<\eta_\ast)=0$,\footnote{This requires $c(\lambda)$ remains
finite in the case when $\lambda\rightarrow1$.} then the scalar
spectrum is divergent, showing the breakdown of the treatment. This
is just what we expect: the theory recovers the general relativity
so early that it looks the same as the usual inflationary model and
the gravitational scalar is not a physical degree of freedom and
will not bring observable power spectrum any more.

\section{PRIMORDIAL GRAVITATIONAL WAVES}
From the discussion above we see that the scalar mode of the
gravitational scalar is scale invariant. This is the generic
property under the de Sitter background with time translation
invariance. We will see the same result for the gravitational tensor
modes\cite{Koh:2009cy}. We also start with the perturbed metric,
only to the tensor parts,
\begin{equation}
ds^{2}=-dt^{2}+a^{2}(t)\left(\delta_{ij}+h_{ij}(t,\vec{x})\right)dx^{i}dx^{j},
\end{equation}
where $h_{ij}$ has been already defined in \ref{gij}, and satisfies
the transverse-traceless conditions $h_{i}^{i}=0$,
$\partial_{i}h_{ij}=0$. Substituting this metric into the total
action, we obtain the tensor action of second order,
\begin{equation}
S_{g}^{(2)}=\int
dtd^{3}xa^{3}\left[\frac{\alpha}{4}\dot{h}_{i}^{j}\dot{h}_{j}^{i}+\beta_{1}\frac{\beta}{4a^{6}}\Delta^{2}h_{i}^{j}\Delta
h_{j}^{i}+\frac{\xi(\lambda)}{4a^{2}}h_{j}^{i}\Delta
h_{i}^{j}\right].\label{eq:s1^2}\end{equation} The transverse
traceless tensor $h_{ij}$ can be Fourier transformed by plane waves
with wavenumber $\mathbf{k}$ as
\begin{equation}
h_{ij}(t,\mathbf{k})=\sum_{A=R,L}\int\frac{d^{3}\mathbf{k}}{(2\pi)^{3}}\psi_{\mathbf{k}}^{A}(t)e^{i\mathbf{k}\cdot\mathbf{x}}p_{ij}^{A},\end{equation}
where $p_{ij}^{A}$ is the circular polarization tensor which is
defined by $ik_{s}\epsilon^{rsj}p_{ij}^{A}=k\rho^{A}p_{i}^{rA}$
\cite{Takahashi:2009wc}. Here $\rho^{R}=1$, $\rho^{L}=-1$, and are
called the right handed mode and the left handed mode respectively.
We also impose normalization conditions as
$p_{j}^{*iA}p_{i}^{jB}=\delta^{AB}$, where $p_{j}^{*iA}$ is the
complex conjugate of $p_{j}^{iA}$. Substituting the expansion into
the action (\ref{eq:s1^2}), we obtain
\begin{equation}
\delta^{2}S_{g}=\sum_{A=R,L}\int
dt\frac{d^{3}\mathbf{k}}{(2\pi)^{3}}a^{3}\left\{\frac{\alpha}{4}\left|\dot{\psi_{\mathbf{k}}^{A}}\right|^{2}-\left[
\beta_{1}\frac{\beta}{4a^{6}}k^{6}+\frac{\zeta(\lambda)}{4a^{2}}\right]
\left|\psi_{\mathbf{k}}^{A}\right|^{2}\right\}.
\end{equation}
Using the variable $\upsilon_{\mathbf{k}}\equiv
a\psi_{\mathbf{k}}^{A}$ and the conformal time $\eta$, and rewrite
the action by co-moving time $\eta$, with $a=-1/H\eta$ in the de
Sitter-like space, we have
\begin{equation}\label{eq:equWAVEsSimple}
 \upsilon^{''}_{\mathbf{k}}(\eta)+\left(k^6H^4\bar{l}^4\eta^4+c^2k^2-\frac{2}{\eta^2}\right)\upsilon_{\mathbf{k}}(\eta)=0.
\end{equation}
where similarly to the treatment on the scalar perturbation before,
we define a characteristic length related to ultraviolet
gravitational waves, $\bar{l}=(\beta_{1}\beta/\alpha)^{1/4}$.
We see from this definition and \eqref{L}, that the characteristic
length of scalar and tensor modes may be different. This relies on
the relative magnitude of different $\beta$'s.

Some discussions parallel to last section will yield the power
spectrum of gravitational wave as
\begin{equation}
\mathscr{P}_{h}=\frac{k^{3}}{2\pi^{2}}\frac{|\upsilon_{\mathbf{k}}|^{2}}{a^{2}}=\left(\frac{H}{2\pi
c}\right)^2_\dag
\end{equation}
where the subscript $\dag$ means the quantities are evaluated at the
time of horizon-crossing of the gravitational waves, i.e.
\begin{equation}
    \frac{a(\eta_\dag)}{k}=\frac{c_{\dag}}{H},\;\;\;\;|\eta_\dag|=\frac{1}{c_{\dag}k}.
\end{equation}
Obviously the power spectrum of tensor mode is scale invariant as
well. In \cite{Koh:2009cy}, the author use the same method to
connect the solution step by step and calculate the infrared power
spectrum, but with only two pieces to join and thus more accurate
sub-solution in each piece: Hankel function in infrared region.
However, after taking the correct connecting time $\eta_\star$ in
\eqref{eta1} as we do before, the dependence on the ``cutoff energy
scale'' there also vanishes.

Now we can calculate the tensor-to-scalar ratio\footnote{Note that
the tensor-to-scalar ratio is usually defined to be
$\mathscr{P}_h/\mathscr{P}_{\delta\phi}$, by the power spectrum of
the perturbations of inflaton field in ordinary inflationary models.
And in super-horizon scales, $\mathscr{P}_{\delta\phi}$ is of the
order as $\mathscr{P}_h$ divided by slow-roll parameter $\epsilon$.
Because we have not placed any scalar field here, we define the
tensor-to-scalar ratio as $\mathscr{P}_h$ divided by just the
spectrum of gravitational scalar},
\begin{equation}\label{ratio}
r=\frac{\mathscr{P}_{h}}{\mathscr{P}_{\psi}}=\frac{c_{s\ast}}{c_\dag}
=\frac{1-\lambda_\ast}{1-3\lambda_\ast}\frac{\xi(\lambda_\ast)}{\xi(\lambda_\dag)}.
\end{equation}
This shows that the ratio is only determined by the speed of
sound/light at the sound-/hubble-horizon-crossing. All the
dependence on the characteristic length of the ultraviolet behavior
do not appear in the ratio. We may try to estimate this ratio by the
assumption that $\xi$ varies slowly to 1 when the $k$-mode we are
interested in crosses the horizons. Then we can neglect the $\xi$
term in (\ref{ratio}) and have only the dependence of $\lambda$ at
sound horizon crossing.  Since $r\ll 1$ from the observations,
$\lambda_\ast=1+2r$ must be very close to $1^+$, which is in
consistency with our former assumption.

\section{Conclusion and Discussion}

In this paper, we clarified several issues in the \hl~ gravity. We
first showed that the strong coupling issue may not be so serious as
argued in the literature before. The basic point is that the
diffeomorphism is only a good approximation even at IR. Taking into
account of the projectability condition, the usual Stuckelberg trick
could not be applied naively. From our discussion, it seems that the
extra dynamical scalar degree of freedom could be decoupled
naturally.

However, the theory may suffer from other pathologies. One concern
is on the existence of the ghost excitation. We showed that as the
perturbations around the flat spacetime, the scalar perturbation
around the flat FRW universe could be a ghost in the parameter
region $\frac{1}{3} < \l < 1$. The presence of the ghost mode is a
serious challenge to the theory. We tried to avoid the dangerous
parameter region by mildly modifying the \hl~ action. We kept only
the most UV sensitive and IR sensitive terms. We discussed the
classical evolution and the power spectra of scalar and tensor
perturbations. We obtained scale invariant spectrum if the Hubble
constant $H$ does not change. We also calculated the tensor-scalar
ratio, and found it could be small under reasonable condition.

The nature of the power spectra studied in this paper is purely
gravitational. In particular, in the language of orthodox cosmology,
the scalar perturbation is expected to set up the initial conditions
and seed the anisotropy of large scalar structure in our universe.
Some work has been done to reveal the evolution of perturbations
after inflation in \hl~gravity\cite{Kobayashi:2009hh}. After
inflation ends, this gravitational perturbation must be converted
into CMB anisotropy and matter inhomogeneity through some
post-inflation evolotions. But still we do not know yet how to
couple the gravitational scalar mode with, for instance, the
radiation. This is an interesting issue, which we would like to
study in future. Recently, an interesting paper on the \hl~universe
with single scalar field discussed the curvature perturbation
$\zeta$\cite{Wang:2009az}.

One essential issue in \hl~ gravity is on its RG flow. In
\cite{Iengo:2009ix}, it has been shown that in Lifshitz-like scalar
field theory, the RG flow may not lead the theory to the fixed point
we want. Considering the numbers of the parameters in modified \hl~
gravity, this raise the concern if the theory can flow to IR fixed
point $\lambda=1$. Moreover, the details of RG flow can tell us if
we can avoid the dangerous region, where the ghost excitation
appears, even we start from a safe region. Furthermore, RG flow may
closely related to the physics in the inflationary era.  It is not
clear whether RG flow of the theory runs to its IR fixed point
before the inflationary era. If it did, then the gravitational
scalar is not dynamical and has nothing to do with inflation. Even
if the energy scale to reach IR limit is lower than the inflation
era, there is an important question to answer: did $\lambda$ vary
significantly in the inflationary era? The variation of $\l$ may
tilt the power spectra and has interesting physical implications. In
any case, the behavior of the \hl~gravity theory under RG flow
deserves careful investigations.

\section*{Acknowledgments}

The work was partially supported by NSFC Grant No.10535060,
10775002, 10975005 and RFDP. BC would like to thank R.G. Cai, Miao
Li and Jian-xin Lu for stimulating discussions. BC also thank
ICTS-USTC for hospitality, where the project was inspired. SP is
grateful to Yi-Fu Cai and Tower Wang for useful directions.

\end{document}